\documentclass[aps,prl,showpacs,onecolumn,superscriptaddress,floatfix]{revtex4-1}
\usepackage{amsmath}
\usepackage{amssymb}
\usepackage{graphicx}
\usepackage{braket}

\begin{document}

\title{Angle-dependent Weiss oscillations in a nanocorrugated two-dimensional electron gas}

\author{Ching-Hao Chang}
\affiliation{Leibniz-Institute for Solid State and Materials Research, Helmholtzstraße 20, 01069 Dresden, Germany}
\email{c.h.chang@ifw-dresden.de}
\author{Carmine Ortix}
\affiliation{Leibniz-Institute for Solid State and Materials Research, Helmholtzstraße 20, 01069 Dresden, Germany}
\affiliation{Institute for Theoretical Physics, Center for Extreme Matter and Emergent Phenomena, Utrecht University, Princetonplein 5, 3584 CC, Utrecht, Netherlands}
\email{c.ortix@uu.nl}

\begin{abstract}
We investigate the diffusive magnetotransport properties of a two-dimensional electron gas residing in a wrinkled nanostructure. The curved geometry of the nanostructure renders an effective inhomogeneous magnetic field which, in turns, yields Weiss oscillations. Since the relative strength of the effective inhomogeneous magnetic field can be tailored by changing the direction of the externally applied magnetic field, these Weiss oscillations exhibit a strong directional dependence. For large external magnetic fields we also find an anisotropic positive magnetoresistance. 
\end{abstract}
\keywords{Weiss oscillations; guiding-center drift; snake orbits; perpendicular magnetic field}
\maketitle 

\section{Introduction}

In very recent years, advanced nanostructuring techniques have enabled the synthesis of novel low-dimensional nanostructures in which thin film materials can be bent into curved deformable objects such as rolled-up nanotubes~\cite{pri00,sch01,rob16}, nanohelices~\cite{zha02}, and even complex nanoarchitectures resembling structures that form naturally in the most basic forms of life~\cite{xu15}. On one hand, these next-generation nanomaterials carry an enormous potential in electronics, ranging from flexible displays to the integration of semiconductor electronics with the soft, curvilinear surfaces of the human body~\cite{rog11,jo15}. On the other hand, the very fundamental quantum mechanical properties of charge carriers in these nanomaterials are strongly affected by the curved background in which they live~\cite{dew57,ort15}. The recent theoretical advances in defining the quantum mechanics in constrained low-dimensional curved geometries have indeed brought to light a series of unique curvature-induced electronic and transport properties. Prime actors include the appearance of winding-generated bound states~\cite{CarmineRUNT}, a strongly anisotropic ballistic magnetoresistance in non-magnetic and spin-orbit-free semiconducting nanotubes~\cite{CHBAMR}, signatures of snaking states in the transport properties of core-shell nanowires~\cite{ros15}, peculiar Aharonov-Bohm oscillations in prismatic core-shell nanowires~\cite{fer09}, the geometric shape control of spin-interference effects in quantum rings~\cite{ber05,yin16}, 
an all-geometrical driving of topological insulating states of matter in bent nanowires~\cite{gen15}, and even reversal of thermoelectric currents in tubular nanowires~\cite{erl17}. 

Among the different geometrically deformed nano-objects that are nowadays produced, rippled nanostructures [c.f. Fig.~\ref{fig:fig1}(a)] undoubtedly play a primary role.
Wrinkling of thin solid films can occur over a wide range of length scales~\cite{wrinklephysics}, and results from the presence of external strains. 
Controlled ripple texturing, for instance, has been achieved in suspended graphene sheets using spontaneously generated or thermally generated strains~\cite{graphenewrinkle}. 
Another route involves the transfer of thin solid films onto pre-stressed elastometers in which strains are suddenly released. This method has enabled the fabrication of wrinkled silicon nanomembranes~\cite{rolledupwrinkle}, and more recently of multilayer black phosphorous flakes with ripple textures~\cite{que16}.
Finally, controlled wrinkling of a nanomembrane with an embedded InGaAs quantum well has been also achieved by partial release and bond back of layers upon underetching~\cite{wrinkleoptic} .  

From the theoretical point of view, the creation of ripples represents an avenue to modulate electronic properties on a local scale with the tremendous advantage that the system remains chemically pristine. Structural ripples can result in added periodic potentials, which can induce mini bandgaps and Shockley end modes~\cite{pan16}. The aim of this paper is to show that the magnetotransport properties of a two-dimensional electron gas (2DEG) in rippled nanostructures naturally display the so-called Weiss oscillations~\cite{Weissosc} occurring in conventional 2DEG when subject to additional periodic electric or magnetic field potentials~\cite{inhomogeneousfield}. This is because when subject to an external homogeneous magnetic field, electrons constrained to a wrinkled structure experience an effective periodically-modulated magnetic field. More importantly, since the ratio between the effective homogeneous and inhomogeneous component of the magnetic field can be tuned on demand by simply rotating the externally applied homogeneous magnetic field, we find the Weiss oscillations to be strongly anisotropic, and consequently appear also in the angular dependence of the magnetoresistance.

%when subject to an external magnetic field, electrons constrained to a wrinkled structure experience an effective periodically-modulated magnetic field. In perfect analogy with the situation encountered in two-dimensional electron gases with additional periodic electric or magnetic field potentials~\cite{inhomogeneousfield}, we thereby find the natural occurrence of Weiss oscillations \cite{Weissosc} in the magnetoresistance. Moreover, since the strength of the effective inhomogeneous field felt by the electrons can be tailored by changing the direction of externally applied magnetic field, these geometrically-induced Weiss oscillations exhibit a strong directional dependence. 

\section{Results and discussion}
\subsection{Quantum theory of a nanocorrugated electron gas with an external magnetic field}
\begin{figure}
\begin{center}
\includegraphics{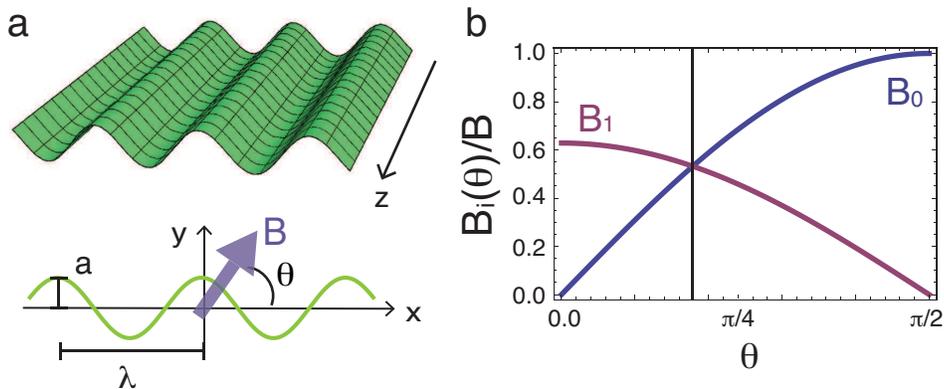}
\end{center}
\caption{(a) Sketch of the surface of a wrinkled nanomembrane (top panel) and the corresponding cross section (bottom panel) . The arrow indicates the externally applied homogeneous magnetic field with its relative direction with respect to the $\hat{x}$ axis. (b) Behavior of the homogeneous ($B_0$) and inhomogeneous ($B_1$) components of the effective magnetic field felt by the electrons as a function of the angle $\theta$ for a nanocorrugated thin film with $a/\lambda=0.1$.}
\label{fig:fig1}
\end{figure}

To investigate the electronic properties of a gas of electrons constrained to a corrugated quasi-two-dimensional nanomembrane, we start out by approximating the corresponding curved surface ${\cal S}$ with a sinusoidal shape whose parametric equations in Euclidean coordinates read  ${\bf r}=\left\{x, y(x), z\right\} \equiv \left\{x, a\cos(2 \pi x/\lambda),z\right\}$, where $a$ represents the maximum wrinkle amplitude while $\lambda$ is its typical wavelength  [see Fig.~\ref{fig:fig1}(a)]. 
To proceed further, we use that the effective two-dimensional Schr\"odinger equation regulating the quantum motion of electrons moving along the curved surface ${\cal S}$ can be derived using the so-called thin-wall quantization procedure \cite{curveSch1}: It treats the quantum motion on a generic curved surface as the limiting case of a quantum particle in three-dimensional space subject to a strong confining force acting in the direction normal to ${\cal S}$. Because of the lateral confinement, quantum 
excitation energies in the normal direction become much higher than in the tangential direction. Henceforth, one can safely ignore the particle motion in the direction normal to the surface, and, on the basis of this, deduce the effective two-dimensional Schr\"odinger equation. This procedure, which is the most rigorous and physically sound one for curved nanosystems embedded in the ordinary Euclidean three-dimensional space, simplifies the problem considerably since the surface curvature is eliminated from the Schr\"odinger equation at the expense of adding to it a potential, dubbed as quantum geometric potential (QGP). The QGP generally cause intriguing phenomena at the nanoscale, such as the emergence of winding-generated bound state in spirally rolled-up nanotubes \cite{CarmineRUNT}. Moreover, in strain-driven nanostructures  the nanoscale variation of strain induced by curvature yields an additional strain-induced geometric potential (SGP) that is of the same functional form of the QGP, but (often strongly) boosting it \cite{strainQGP}. 
With both the QGP and SGP taken into account, the effective Hamiltonian for a nanocorrugated electron gas takes a form, which, after  employing the translational invariance along the $\hat{z}$ direction and thereby separating the two-dimensional wavefunction as $\Psi (x,z)=\psi (x)\times e^{i k_z z}$, reads 
\begin{align}
{\cal H}^{0}= -\frac{\hbar^{2} }{2m^{\star}}\frac{\partial_{x}}{\sqrt{g_{x,x}}}\bigg(\frac{\partial_{x}}{\sqrt{g_{x,x}}}\bigg)+v_{R}{\cal V}_{G}
+\frac{\hbar^2k_z^2}{2m^{\star}}.
\label{eq:h0}
\end{align} 
In the equation above,  $m^\star$ is the electron effective mass, $g_{x,x} = 1+[\partial_{x}y(x)]^2$ is the only non-trivial metric tensor component of the curved surface ${\cal S}$, 
whereas the functional form of the geometric potential is 
\begin{align}
{\cal V}_{G}=-\frac{\hbar^2}{8 m^\star g^3_{x,x}}\left[\partial^2_x y(x)\right]^2. 
\end{align}
The prefactor $v_{R}$ characterizing the strength of the geometric potential in Eq.~\ref{eq:h0} measures the renormalization of the QGP due to strain effects. It is therefore a material-dependent parameter, which can be estimated to be $v_R=6$ in GaAs heterostructures, and $v_R=600$ in Si inversion layers assuming a 2 nm effective thickness  \cite{reviewBAMR}. 

It has been recently shown that the thin-wall quantization procedure can be safely formulated even in the presence of externally applied electric and magnetic fields \cite{curveSch2,curveSch3}, which allows the investigation of unique curvature-induced effects on  the magnetotransport properties of a two-dimensional electron gas\cite{reviewBAMR,CHBAMR}. We therefore consider the effect of an externally applied homogeneous magnetic field, which is orthogonal to the translationally invariant direction of the nanomembrane, and forms an angle $\theta$ with the $\hat{x}$ direction [see Fig.~\ref{fig:fig1}(a)], {\it i.e.} ${\bf B}=B \left\{\cos\theta,\sin\theta,0 \right\}$. 
We next fix the gauge in which the corresponding vector potential reads:  ${\bf  A}=\left\{0,0,A_z(\theta)\right\}$ where $A_z(\theta)=B(y \cos\theta-x\sin\theta)$. Since in this gauge the vector potential does not have a finite component in the direction normal to the curved surface ${\cal S}$, we can straightforwardly apply the framework of Ref.\cite{curveSch2,curveSch3}, thereby obtaining the effective two-dimensional Hamiltonian: 
\begin{align}
{\cal H} = -\frac{\hbar^{2} }{2m^{\star}}\frac{\partial_{x}}{\sqrt{g_{x,x}}}\bigg(\frac{\partial_{x}}{\sqrt{g_{x,x}}}\bigg)
+\frac{\hbar^2}{2m^{\star}}\Big(k_z-eA_z(\theta)\Big)^2+v_{R}{\cal V}_{G}.
\label{eq:h}
\end{align} 
The Hamiltonian in the equation above can be interpreted as the effective one-dimensional Hamiltonian regulating the quantum motion of a particle constrained to a serpentine-shaped waveguide in the presence of both an attractive potential generated by the geometric curvature, and a $k_z$-dependent magnetic potential.  Although both these two potential should be treated on an equal footing, we note that for corrugated nanomembranes with typical wavelengths of the order of a few hundreds of nanometers the geometric potential can be safely disregarded  for $v_R \simeq 10^0$. This is because the Landau magnetic length ($\approx 26/\sqrt{B{\rm (tesla)}}$ nm) corresponding to an external magnetic field of a few teslas is much smaller than the corrugation period, thereby implying that the amplitude of the QGP is more than two orders of magnitude smaller than the magnetic potential. In the remainder of this article, we will consider a corrugated two-dimensional electron gases formed at GaAs-based heterostructures where, as mentioned above, $v_R \simeq 6$, and therefore the last term in Eq.~\ref{eq:h} can be neglected. 

To make further progress, we rewrite the effective Hamiltonian by introducing the arclength of the sinusoidal profile measured from $x=0$, $s=\int_0^{x} \sqrt{g_{x,x}}~dx$ \cite{CarmineRUNT}, in terms of which Eq.~\ref{eq:h} can be recast as 
\begin{align}
{\cal H} = -\frac{\hbar^{2} }{2m^{\star}} \partial^2_s +\frac{\hbar^2}{2m^{\star}}\Big(k_z-eA_z(\theta)\Big)^2,
\label{eq:heff}
\end{align}  
with 
\begin{align}
A_z(\theta) \approx B \left[a\cos{\dfrac{2 \pi s}{\lambda}} \cos\theta-s\sin\theta\right]. 
\label{eq:effA}
\end{align}
In the equation above, we have used that in the shallow deformation limit $a/\lambda \ll 1$, $s \approx x$. Most importantly, Eqs.~\ref{eq:heff},\ref{eq:effA} correspond to the Hamiltonian in the effective mass approximation for a two-dimensional electron gas subject to a perpendicular magnetic field  
\begin{align}
B(s)= B \left[\dfrac{2 \pi a}{\lambda}\sin{\dfrac{2\pi s}{\lambda}} \cos{\theta}+\sin\theta \right].
\label{eq:effB}
\end{align}
For a given angle $\theta$, the effective field in Eq.(\ref{eq:effB}) has an homogeneous component of strength $B_0 = B \sin\theta$, as well as an inhomogenous component  $B_1 = (2\pi  B a/\lambda)  \cos{\theta}$. As a result, depending on the direction of the externally applied homogeneous magnetic field, the effective field felt by electrons constrained to the corrugated nanomembrane changes from being purely homogeneous for $\theta=\pi/2$ to a zero-average periodic magnetic field for $\theta=0$ [c.f. Fig.~\ref{fig:fig1}(b)]. In the next section, we will analyze the classical magnetic trajectories \cite{CHCAMR} that are thereby realized by rotating the external magnetic field.

\subsection{Classical electron trajectories}

\begin{figure}
\begin{center}
\includegraphics[width=.75\columnwidth]{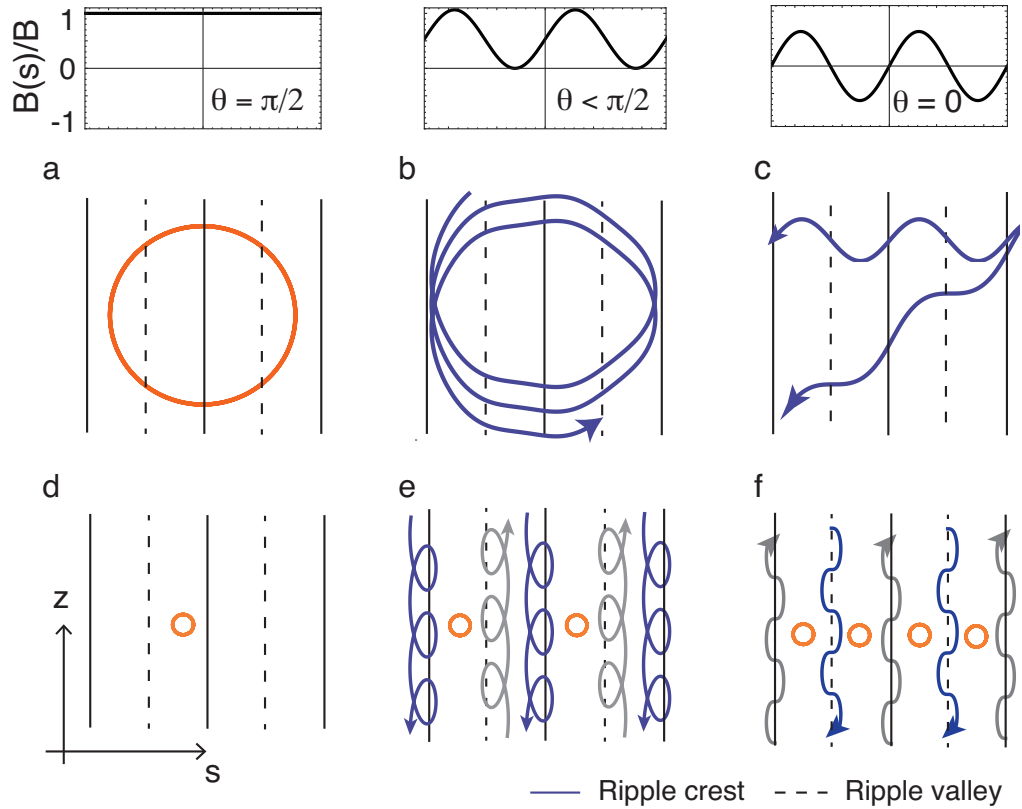}
\caption{Top panels: Effective magnetic fields felt by the electrons in a corrugated two-dimensional electron gas as obtained by changing the direction of an externally applied magnetic field. Middle panels: In the weak-field regime the usual cyclotron orbits obtained for a magnetic field direction $\theta=\pi/2$ (a) acquire a guiding-center drift which is reinforced under resonance condition (b). For the magnetic field direction $\theta=0$ the absence of an homogeneous component of the effective magnetic field yields instead slightly perturbed transverse orbits (c). Bottom panels: in the strong-field regime small radii cyclotron orbits (d) can coexist with drifting orbits (e) and snake states with reversed direction (f) localized at the crests and valleys of the wrinkles in the presence of an inhomogeneous component of the effective magnetic field.} 
\label{fig:fig2}
\end{center}
\end{figure}

We first analyze the classical electron trajectories  in a nanocorrugated two-dimensional electron gas subject to an homogeneous external magnetic field. 
Since, as demonstrated above, we can switch from the native curved system to a flat one in which the effective magnetic field both possesses an homogenous and an inhomogeneous component, the electron trajectories can be obtained, in perfect analogy with Ref.~\cite{zwe99}, by solving the classical equation of motion  $m^\star \dot{{\bf v}}=-e~{\bf v}\times{\bf B}(s)$ with ${\bf B}(s)$ given in Eq.~\ref{eq:effB}.
When the inhomogeneous component vanishes, {\it i.e.} assuming an external magnetic field that points along the $\hat{y}$ axis ($\theta=\pi/2$) in Fig.~\ref{fig:fig1}(a), 
the classical trajectories simply corresponds to cyclotron orbits with a radius  $R_{\rm cycl}= m^\star v/(e B_0)$ that is inversely proportional to the magnetic field strength  [see Figs.\ref{fig:fig2}(a),(d)]. 

By rotating the magnetic field away from $\hat{y}$ direction, the finite inhomogeneous component of the effective magnetic field leads to two different types of characteristic electron orbits depending on whether or not the angle-dependent mean cyclotron radius $\bar{R}_{\rm cycl}= m^\star v/(e B_0 \sin{\theta})$ exceeds the wrinkle wavelength $\lambda$. 
In the weak-field regime, {\it i.e.} for $\bar{R}_{\rm cycl}>\lambda$, the inhomogeneous component of the magnetic field yields a drift of the electron orbit center which is simply superimposed to the unperturbed cyclotron motion due to the homogeneous part of the magnetic field. In complete analogy with the situation encountered in a two-dimensional electron gas with an electrostatic potential grating~\cite{Weisstheory,Weisstheory2}, this guiding-center drift
is oscillatory and can be neglected except when it resonates with the periodic cyclotron motion [see Fig.~\ref{fig:fig2}(b)]. 
In the $R_{\rm cycl}<\lambda$ regime, instead, the homogeneous component of the magnetic field is strong enough to localize the electrons in different space regions of the wrinkles. More specifically, close to the regions where the gradient of the effective magnetic field is zero, and thus the local curvature of the wrinkle vanishes,  the electronic trajectories form usual cyclotron orbits whose radius depends on the local magnetic field strength. Close to the valleys and crests of the wrinkles, instead, the gradient of the effective magnetic field is maximal and thus the cyclotron orbits acquire a strong drift along the $\hat{z}$ and the $-\hat{z}$ direction respectively [see Fig.~\ref{fig:fig2}(e)].

When the externally applied homogeneous magnetic field points exactly towards the $\hat{x}$ direction in Fig.~\ref{fig:fig1}(a) the situation changes drastically. The homogeneous component of the effective magnetic field identically vanishes and thus in the weak-field regime, the electron orbits corresponds to slightly perturbed transverse orbits  [see Fig.~\ref{fig:fig2}(c)].  Cranking up the strength of the magnetic field to the regime in which the characteristic cyclotron radius $\widetilde{R}_{\rm cycl}= m^\star v/(e B_1)$ is smaller than the wrinkle wavelength yields electronic orbits strongly resembling the ones encountered for $\theta < \pi/2$ but with the following caveat: since the effective magnetic field averages to zero, the drifting cyclotron orbits of Fig.~\ref{fig:fig2}(e) are substituted by snake states with opposite velocity similarly 
to the situation encountered in tubular nanostructures subject to a transversal magnetic field  \cite{ros15,bel10}. 

\subsection{Angle-dependent Weiss oscillations}
Having established the nature of the classical electron trajectories, we now analyze their impact on the magnetotransport properties of a nanocorrugated two-dimensional electron gas subject to an externally applied magnetic field.  
To do so, we first solve the magnetic spectra by diagonalizing the effective Hamiltonian Eq.(\ref{eq:heff}) using as complete basis the Landau levels states due to the homogeneous field $B_0$ \cite{diagonal1,diagonal2}.
\begin{figure}
\begin{center}
\includegraphics[width=.65\columnwidth]{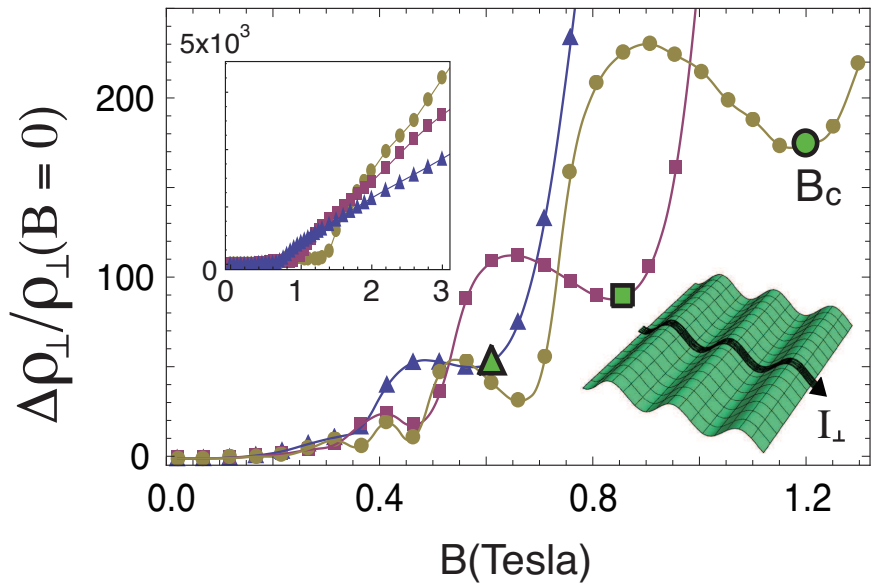}
\caption{Weiss oscillation in the magnetoresistance of a nanocorrugated two-dimensional electron gas subject to an homogenous external magnetic field. 
We have considered the amplitude and period of the corrugations to be $a=25~$nm and $\lambda=250~$nm respectively. The direction of the external magnetic field forms an angle $\theta \simeq 30^{\circ}$ with the $\hat{x}$ axis [c.f. Fig.~\ref{fig:fig1}(a)]. The magnetoresistance is computed for three carrier densities corresponding to $n=0.5,1,2\times10^{12}$~cm$^{-2}$, which are denoted by triangles, squares and disks, respectively. We also show the critical fields at which the Weiss oscillations are substituted by the onset of a positive magnetoresistance [see the inset]. The schematic figure shows the direction of the current along the wrinkled nanostructure.}
\label{fig:fig3}
\end{center}
\end{figure}
Thereafter, we compute the dc conductivity reading~\cite{WeissSdH}
\begin{align}
\sigma_{i j}=i\frac{\hbar e^2}{A}\sum_{\xi,\xi'}\int dE\int dE' P(E-E_{\xi})P(E'-E_{\xi'})\times (f_E-f_{E'})\frac{\langle\xi|v_i|\xi'\rangle\langle\xi'|v_j|\xi\rangle}{(E-E')^2},
\label{eq:con}
\end{align}
where $i,j={s,z}$, $\ket{\xi}=\ket{n,k_z}$ is an eigenstate with subband index $n$ and longitudinal momentum $k_z$, 
$A$ is the area of the sample, $v_i$ is the group velocity operator, and $f_E$ is the Fermi-Dirac distribution function. In order to explicitly take into account the impurity broadening effect on the conductivity, which is important at low temperatures, we introduced an impurity broadening $P(E)=(\Gamma/\pi)/(E^2+\Gamma^2)$ with a constant width $\Gamma = \hbar/\tau$, $\tau$ being a constant relaxation time. 
Eq.~\ref{eq:con} explicitly accounts for both the diagonal and the non-diagonal diffusion contributions to the dc conduction but disregards the collisional conductivity describing transport through localized states due to impurity scattering. The latter, however, is  mainly responsible for the Shubnikov-de Haas (SdH) oscillations~\cite{WeissSdH}, and gives only a subleading contribution for the occurrence of the Weiss oscillations  investigated in this study. Therefore, we will only account for the diffusional part of the dc conduction in  
our system. We emphasize that previous works on conventional 2DEG subject to additional modulated magnetic fields have accounted only for the diagonal diffusion contribution to the dc conductance~\cite{diagonal1,diagonal2,Peeters1}. This is a well justified assumption considering a magnetic modulation weak as compared to the external homogeneous magnetic field. In our case, instead, the off-diagonal contribution is essential in the small $\theta$ regime, where the 2DEG is in a lateral magnetic superlattice~\cite{Peeters2} perturbed by a weak homogeneous magnetic field.

To obtain the conductivity tensor components from Eq.~\ref{eq:con}, we fix the temperature to $T=3.66$K, use the relaxation time $\tau=3.66 \times 10^{-11}$s~\cite{WeissSdH}, and introduce a cutoff in the subband index $n=200$, thereby reaching an accuracy within 1$\%$. The resistivity tensor is thereafter found by inverting the conductivity tensor. 
Fig.~\ref{fig:fig3} shows the ensuing behavior of the magnetoresistance (MR) in the direction orthogonal to the wrinkles $\Delta \rho_{\perp}/\rho_{\perp}(B=0)= \rho_{ss}(B)/\rho_{ss}(0)-1$ assuming the homogeneous external magnetic field forms an angle $\theta \simeq 30^{\circ}$ with respect to the $\hat{x}$ direction [c.f. Fig.~\ref{fig:fig1}(a)]. 
The presence of both an homogeneous and an inhomogeneous component of the effective magnetic field felt by the electron gas leads, independent of the carrier density, to the presence of Weiss oscillations, which are the prime physical consequence of the guiding-center drift resonance discussed above.  
These Weiss oscillations, however, are clearly visible for magnetic fields smaller than a critical value $B_c \propto \sqrt{n}$ with $n$ the carrier density [see Fig.~\ref{fig:fig3}]. 
In the $B>B_c$ regime, indeed, we observe a strong positive diffusive magnetoresistance [c.f. the inset of Fig.~\ref{fig:fig3}], which is due to the fact that the inhomogeneous component of the effective magnetic field is comparable in magnitude to the homogeneous one. We point out that this feature is unique of our system and is not generally realized in conventional two-dimensional electron gas with additional ferromagnetic or superconducting strips where the characteristic strength of the modulated magnetic field is much smaller than the homogeneous one~\cite{WeissSdH}.

\begin{figure}
\begin{center}
\includegraphics{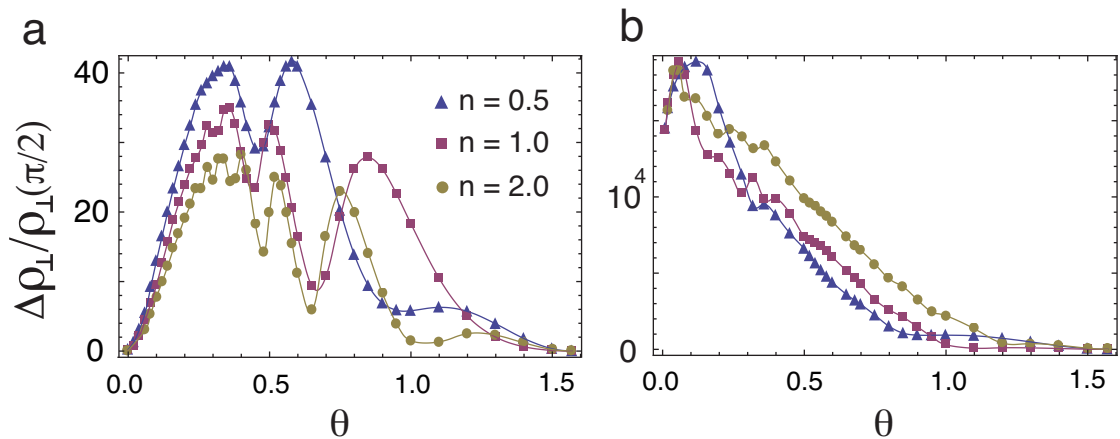}
\caption{Angle dependence of the magnetoresistance  $\rho_{ss}(\theta) / \rho_{ss}(\pi/2) - 1$ in a nanocorrugated electron gas subject to an homogeneous external magnetic field with strength $B=0.4$~T (a) and $B=4$T (b) for three carrier densities corresponding to $n=0.5,1,2\times 10^{12}$~cm$^{-2}$.}
\label{fig:fig4}
\end{center}
\end{figure}

%\begin{align}
%B_c \approx 0.8 \frac{m^\star v_F}{e \lambda \sin\theta} = 0.8 \frac{\hbar k_F}{e \lambda \sin\theta} \propto n^{1/2},
%\label{eq:bc}
%\end {align}
%where $v_F$ is the Fermi velocity, $k_F = \sqrt{2\pi n}$ is the Fermi wavevector, and $n = k^2_F/(2\pi)$ is the carrier density of GaAs 2DEGs \cite{be-review}.
%The value in Eq.(\ref{eq:bc}) refers to that a cyclotron radius at Fermi level saturates the wrinkle wavelength and $0.8$ is a prefactor obtained numerically by the results in Fig.\ref{fig:fig3}. 

The tunability of the ratio between the effective inhomogeneous magnetic field strength and the homogeneous one, as obtained by simply rotating the direction of the external magnetic field, also yields a strong directional dependence of the magnetotransport properties. This is demonstrated in Fig.~\ref{fig:fig4}(a,b) where we show the angle dependence of the magnetoresistance $\Delta \rho_{\perp} / \rho_{\perp}(\pi/2) \equiv \rho_{ss}(\theta) / \rho_{ss}(\pi/2) - 1$. For very-weak fields [c.f. Fig.~\ref{fig:fig4}(a)], the directional dependence of the magnetoresistance displays Weiss oscillations, which result from the angle-dependence of the effective field homogeneous component $B \sin{\theta}$. Moreover, we find that the longitudinal magnetoresistance for $\theta=0$ and $\theta=\pi/2$, {\it i.e.} for an external magnetic field pointing along the $\hat{x}$ and $\hat{y}$ direction in Fig.~\ref{fig:fig1}(a), is practically the same. This is a consequence of the fact that for these two particular directions the classical electron trajectories correspond to the usual cyclotron and transverse orbits, which yield an equal diffusive magnetoresistance at the classical level. 
In the strong-field regime instead, [c.f. Fig.~\ref{fig:fig4}(b)], the occurrence of Weiss oscillations in the magnetoresistance angle dependence  is precluded by the onset of the strong positive magnetoresistance discussed above. However, we still find a marked anisotropy with the magnetoresistance that becomes very strong when the inhomogeneous component of the effective magnetic field dominates over the homogeneous one, thereby suggesting a leading contribution due to the appearance of snake states [c.f. Fig.~\ref{fig:fig2}].  
 
\subsection{Discussion}

To conclude, we have shown that the geometry of a nanocorrugated two-dimensional electron gas can be used to engineer an effective periodically inhomogeneous magnetic field coexisting with an homogeneous one by simply applying a constant external magnetic field. As such, our proposed setup does not necessitate the usage of ferromagnetic or superconducting strips as routinely done in conventional semiconducting heterostructures. The magnetotransport properties of these patterned nanostructures have been extensively theoretically studied~\cite{diagonal1,diagonal2,Peeters1} in the regime of weak magnetic modulations. The geometrical engineering of modulated magnetic fields presented in this work, instead,
can be conveniently used to access regimes where the modulated magnetic field 
is comparable in strength to the homogeneous one. Such a regime can be also realized in conventional flat nanostructures using magnetic barriers. However, in the latter situation the step-like character of the magnetic field inhomogeneity leads to localized charge densities~\cite{Peeters3}, which are not expected to occur for the smooth magnetic field profile realized in our setup. Finally our setup also allows to mimic a lateral magnetic superlattice~\cite{Peeters2,car95} perturbed by an homogeneous magnetic field with tunable strength (tilting the magnetic field towards the ${\hat y}$ direction in Fig.~\ref{fig:fig1}(a)). The quadratic positive magnetoresistance we find in this situation [c.f.~Fig.\ref{fig:fig4}(a)] agrees with experimental observations~\cite{InhomoB1}.
We expect that the angular dependence of the Weiss oscillations at weak magnetic field and the strong anisotropic positive magnetoresistance at strong external fields predicted in our study can be not only tested in wrinkled nanomembranes with embedded quantum wells as obtained by partial release and bond back of layers by selective underetching ~\cite{wrinkleoptic}, but also in two-dimensional materials where ripples can be strain-engineered by depositing atomic crystals on prestreched elastomeric substrates~\cite{que16}.

\section{Acknowledgements}
We acknowledge the financial support of the Future and Emerging Technologies (FET) programme within
the Seventh Framework Programme for Research of the European Commission 
under FET-Open grant number: 618083 (CNTQC).  
C.O. acknowledges support from the Deutsche Forschungsgemeinschaft (Grant No. OR 404/1-1), and from a VIDI grant (Project 680-47-543) financed by the Netherlands Organization for Scientific Research (NWO). 

\section*{References}

\providecommand{\newblock}{}

%\bibliographystyle{iopart-num}
%\bibliography{wrinkle,nanomat,matsciexp,matscitheo}

\begin{thebibliography}{10}
\expandafter\ifx\csname url\endcsname\relax
  \def\url#1{{\tt #1}}\fi
\expandafter\ifx\csname urlprefix\endcsname\relax\def\urlprefix{URL }\fi
\providecommand{\eprint}[2][]{\url{#2}}
% Bibliography created with iopart-num v2.1
% /biblio/bibtex/contrib/iopart-num

\bibitem{pri00}
Prinz V~Y, Seleznev V~A, Gutakovsky A~K, Chehovskiy A~V, Preobrazhenskii V~V,
  Putyato M~A and Gavrilova T~A 2000 {\em Physica E (Amsterdam)\/} {\bf 6} 828
  -- 831 
  %ISSN 1386-9477
  %  \urlprefix\url{http://www.sciencedirect.com/science/article/B6VMT-3YSY07V-70/2/64d79d338fd8344816bdc8c0d38d6b2a}

\bibitem{sch01}
{Schmidt} O~G and {Eberl} K 2001 {\em Nature\/} {\bf 410} 168

\bibitem{rob16}
Streubel R, Fischer P, Kronast F, Kravchuk V~P, Sheka D~D, Gaididei Y, Schmidt
  O~G and Makarov D 2016 {\em Journal of Physics D: Applied Physics\/} {\bf 49}
  363001 
  %\urlprefix\url{http://stacks.iop.org/0022-3727/49/i=36/a=363001}

\bibitem{zha02}
Zhang H~F, Wang C~M and Wang L~S 2002 {\em Nano Letters\/} {\bf 2} 941

\bibitem{xu15}
Xu S, Yan Z, Jang K~I, Huang W, Fu H, Kim J, Wei Z, Flavin M, McCracken J, Wang
  R, Badea A, Liu Y, Xiao D, Zhou G, Lee J, Chung H~U, Cheng H, Ren W, Banks A,
  Li X, Paik U, Nuzzo R~G, Huang Y, Zhang Y and Rogers J~A 2015 {\em Science\/}
  {\bf 347} 154--159 
  %(\textit{Preprint}
 % \eprint{http://www.sciencemag.org/content/347/6218/154.full.pdf})
 % \urlprefix\url{http://www.sciencemag.org/content/347/6218/154.abstract}

\bibitem{rog11}
Rogers J~A, Lagally M~G and Nuzzo R~G 2011 {\em Nature (London)\/} {\bf 477} 45

\bibitem{jo15}
Jo J~W, Kim J, Kim K~T, Kang J~G, Kim M~G, Kim K~H, Ko H, Kim Y~H and Park S~K
  2015 {\em Advanced Materials\/} {\bf 27} 1182--1188 
  %ISSN 1521-4095
  %\urlprefix\url{http://dx.doi.org/10.1002/adma.201404296}

\bibitem{dew57}
DeWitt B~S 1957 {\em Rev.\ Mod.\ Phys.\/} {\bf 29} 377

\bibitem{ort15}
Ortix C 2015 {\em Phys. Rev. B\/} {\bf 91}(24) 245412
 % \urlprefix\url{https://link.aps.org/doi/10.1103/PhysRevB.91.245412}

\bibitem{CarmineRUNT}
Ortix C and van~den Brink J 2010 {\em Phys. Rev. B\/} {\bf 81}(16) 165419
  %\urlprefix\url{https://link.aps.org/doi/10.1103/PhysRevB.81.165419}

\bibitem{CHBAMR}
Chang C~H, van~den Brink J and Ortix C 2014 {\em Phys. Rev. Lett.\/} {\bf
  113}(22) 227205
  %\urlprefix\url{https://link.aps.org/doi/10.1103/PhysRevLett.113.227205}

\bibitem{ros15}
Rosdahl T~O, Manolescu A and Gudmundsson V 2015 {\em Nano Letters\/} {\bf 15}
  254--258 
  %pMID: 25426964 
  %(\textit{Preprint}
  %\eprint{http://dx.doi.org/10.1021/nl503499w})
  %\urlprefix\url{http://dx.doi.org/10.1021/nl503499w}

\bibitem{fer09}
Ferrari G, Goldoni G, Bertoni A, Cuoghi G and Molinari E 2009 {\em Nano
  Letters\/} {\bf 9} 1631--1635 
  %(\textit{Preprint}
  %\eprint{http://dx.doi.org/10.1021/nl803942p})
  %\urlprefix\url{http://dx.doi.org/10.1021/nl803942p}

\bibitem{ber05}
Bercioux D, Frustaglia D and Governale M 2005 {\em Phys. Rev. B\/} {\bf 72}(11)
  113310 
  %\urlprefix\url{https://link.aps.org/doi/10.1103/PhysRevB.72.113310}

\bibitem{yin16}
Ying Z~J, Gentile P, Ortix C and Cuoco M 2016 {\em Phys. Rev. B\/} {\bf 94}(8)
  081406 
  %\urlprefix\url{https://link.aps.org/doi/10.1103/PhysRevB.94.081406}

\bibitem{gen15}
Gentile P, Cuoco M and Ortix C 2015 {\em Phys. Rev. Lett.\/} {\bf 115}(25)
  256801
  %\urlprefix\url{https://link.aps.org/doi/10.1103/PhysRevLett.115.256801}

\bibitem{erl17}
Erlingsson S~I, Manolescu A, Nemnes G~A, Bardarson J~H and Sanchez D 2017 {\em
  Phys. Rev. Lett.\/} {\bf 119}(3) 036804
  %\urlprefix\url{https://link.aps.org/doi/10.1103/PhysRevLett.119.036804}

\bibitem{wrinklephysics}
Cerda E and Mahadevan L 2003 {\em Phys. Rev. Lett.\/} {\bf 90}(7) 074302
  %\urlprefix\url{https://link.aps.org/doi/10.1103/PhysRevLett.90.074302}

\bibitem{graphenewrinkle}
Bao W, Miao F, Chen Z, Zhang H, Jang W, Dames C and Lau C~N 2009 {\em Nat
  Nano\/} {\bf 4} 562--566

\bibitem{rolledupwrinkle}
Guo Q, Zhang M, Xue Z, Ye L, Wang G, Huang G, Mei Y, Wang X and Di Z 2013 {\em
  Applied Physics Letters\/} {\bf 103} 264102 
  %(\textit{Preprint}
  %\eprint{http://dx.doi.org/10.1063/1.4857875})
  %\urlprefix\url{http://dx.doi.org/10.1063/1.4857875}

\bibitem{que16}
Quereda J, San-Jose P, Parente V, Vaquero-Garzon L, Molina-Mendoza A~J, Agraït
  N, Rubio-Bollinger G, Guinea F, Roldán R and Castellanos-Gomez A 2016 {\em
  Nano Letters\/} {\bf 16} 2931--2937 
  %pMID: 27042865 (\textit{Preprint}
  %\eprint{http://dx.doi.org/10.1021/acs.nanolett.5b04670})
  %\urlprefix\url{http://dx.doi.org/10.1021/acs.nanolett.5b04670}

\bibitem{wrinkleoptic}
Mei Y, Kiravittaya S, Benyoucef M, Thurmer D~J, Zander T, Deneke C, Cavallo F,
  Rastelli A and Schmidt O~G 2007 {\em Nano Letters\/} {\bf 7} 1676--1679 
  %pMID:
  %17461606 (\textit{Preprint} \eprint{http://dx.doi.org/10.1021/nl070653e})
  %\urlprefix\url{http://dx.doi.org/10.1021/nl070653e}

\bibitem{pan16}
Pandey S and Ortix C 2016 {\em Phys. Rev. B\/} {\bf 93}(19) 195420
  %\urlprefix\url{https://link.aps.org/doi/10.1103/PhysRevB.93.195420}

\bibitem{Weissosc}
Ye P~D, Weiss D, Gerhardts R~R, Seeger M, von Klitzing K, Eberl K and Nickel H
  1995 {\em Phys. Rev. Lett.\/} {\bf 74}(15) 3013--3016
  %\urlprefix\url{https://link.aps.org/doi/10.1103/PhysRevLett.74.3013}

\bibitem{inhomogeneousfield}
Nogaret A 2010 {\em Journal of Physics: Condensed Matter\/} {\bf 22} 253201
%  \urlprefix\url{http://stacks.iop.org/0953-8984/22/i=25/a=253201}

\bibitem{curveSch1}
da~Costa R~C~T 1981 {\em Phys. Rev. A\/} {\bf 23}(4) 1982--1987
  %\urlprefix\url{https://link.aps.org/doi/10.1103/PhysRevA.23.1982}

\bibitem{strainQGP}
Ortix C, Kiravittaya S, Schmidt O~G and van~den Brink J 2011 {\em Phys. Rev.
  B\/} {\bf 84}(4) 045438
  %\urlprefix\url{https://link.aps.org/doi/10.1103/PhysRevB.84.045438}

\bibitem{reviewBAMR}
Chang C~H and Ortix C 2017 {\em International Journal of Modern Physics B\/}
  {\bf 31} 1630016 
  %(\textit{Preprint}
  %\eprint{http://www.worldscientific.com/doi/pdf/10.1142/S0217979216300164})
  %\urlprefix\url{http://www.worldscientific.com/doi/abs/10.1142/S0217979216300164}

\bibitem{curveSch2}
Ferrari G and Cuoghi G 2008 {\em Phys. Rev. Lett.\/} {\bf 100}(23) 230403
  %\urlprefix\url{https://link.aps.org/doi/10.1103/PhysRevLett.100.230403}

\bibitem{curveSch3}
Ortix C and van~den Brink J 2011 {\em Phys. Rev. B\/} {\bf 83}(11) 113406
  %\urlprefix\url{https://link.aps.org/doi/10.1103/PhysRevB.83.113406}

\bibitem{CHCAMR}
Chang C~H and Ortix C 2017 {\em Nano Letters\/} {\bf 17} 3076--3080 
%pMID:  28394625
  % (\textit{Preprint}
 % \eprint{http://dx.doi.org/10.1021/acs.nanolett.7b00426})
  %\urlprefix\url{http://dx.doi.org/10.1021/acs.nanolett.7b00426}

\bibitem{zwe99}
Zwerschke S~D~M, Manolescu A and Gerhardts R~R 1999 {\em Phys. Rev. B\/} {\bf
  60}(8) 5536--5548
  %\urlprefix\url{https://link.aps.org/doi/10.1103/PhysRevB.60.5536}

\bibitem{Weisstheory}
Beenakker C~W~J 1989 {\em Phys. Rev. Lett.\/} {\bf 62}(17) 2020--2023
  %\urlprefix\url{https://link.aps.org/doi/10.1103/PhysRevLett.62.2020}

\bibitem{Weisstheory2}
Peeters F~M and Vasilopoulos P 1992 {\em Phys. Rev. B\/} {\bf 46}(8) 4667--4680
  %\urlprefix\url{https://link.aps.org/doi/10.1103/PhysRevB.46.4667}

\bibitem{bel10}
Bellucci S and Onorato P 2010 {\em Phys. Rev. B\/} {\bf 82}(20) 205305
  %\urlprefix\url{https://link.aps.org/doi/10.1103/PhysRevB.82.205305}

\bibitem{diagonal1}
Wu X and Ulloa S~E 1993 {\em Phys. Rev. B\/} {\bf 47}(12) 7182--7186
 % \urlprefix\url{https://link.aps.org/doi/10.1103/PhysRevB.47.7182}

\bibitem{diagonal2}
Shi Q~W and Szeto K~Y 1997 {\em Phys. Rev. B\/} {\bf 55}(7) 4558--4562
  %\urlprefix\url{https://link.aps.org/doi/10.1103/PhysRevB.55.4558}

\bibitem{WeissSdH}
Shi J, Peeters F~M, Edmonds K~W and Gallagher B~L 2002 {\em Phys. Rev. B\/}
  {\bf 66}(3) 035328
  %\urlprefix\url{https://link.aps.org/doi/10.1103/PhysRevB.66.035328}

\bibitem{Peeters1}
Peeters F~M and Vasilopoulos P 1993 {\em Phys. Rev. B\/} {\bf 47}(3) 1466--1473
  %\urlprefix\url{https://link.aps.org/doi/10.1103/PhysRevB.47.1466}

\bibitem{Peeters2}
Ibrahim I~S and Peeters F~M 1995 {\em Phys. Rev. B\/} {\bf 52}(24) 17321--17334
  %\urlprefix\url{https://link.aps.org/doi/10.1103/PhysRevB.52.17321}

\bibitem{Peeters3}
Ibrahim I~S, Schweigert V~A and Peeters F~M 1997 {\em Phys. Rev. B\/} {\bf
  56}(12) 7508--7516
  %\urlprefix\url{https://link.aps.org/doi/10.1103/PhysRevB.56.7508}

\bibitem{car95}
Carmona H~A, Geim A~K, Nogaret A, Main P~C, Foster T~J, Henini M, Beaumont S~P
  and Blamire M~G 1995 {\em Phys. Rev. Lett.\/} {\bf 74}(15) 3009--3012
  %\urlprefix\url{https://link.aps.org/doi/10.1103/PhysRevLett.74.3009}

\bibitem{InhomoB1}
Nogaret A, Carlton S, Gallagher B~L, Main P~C, Henini M, Wirtz R, Newbury R,
  Howson M~A and Beaumont S~P 1997 {\em Phys. Rev. B\/} {\bf 55}(24)
  R16037--R16040
  %\urlprefix\url{https://link.aps.org/doi/10.1103/PhysRevB.55.R16037}

\end{thebibliography}

\end{document}